\documentclass[12pt]{article}
\usepackage{cite}
\usepackage{color}
\usepackage{graphicx}
\usepackage{amsmath}
\usepackage{amssymb}
\usepackage{xspace}

\def\baselinestretch{1.2}
\newcommand{\be}{\begin{equation}}
\newcommand{\ee}{\end{equation}}
\newcommand{\beq}{\begin{eqnarray}}
\newcommand{\eeq}{\end{eqnarray}}

\newcommand{\gone}[1]{{}}
\newcommand{\bl}{\noindent $\bullet\ $}

\begin{document}
\begin{titlepage}
\begin{flushright}
MAD-TH-08-03
\end{flushright}

\vfil\

\begin{center}

{\Large{\bf Microscopic Formulation of Puff Field Theory}}

\vfil

Sheikh Shajidul Haque and Akikazu Hashimoto

\vfil

Department of Physics, University of Wisconsin, Madison, WI 53706

\vfil

\end{center}

\begin{abstract}
\noindent We describe a generalization of Puff Field Theory to $p+1$ dimensions where $0 \le p \le 5$. We then focus on the case of $p=0$, ``Puff Quantum Mechanics,'' and construct a formulation independent of string theory. 
\end{abstract}
\vspace{0.5in}

\end{titlepage}
\renewcommand{\baselinestretch}{1.05}  

Melvin twist, also known as the T-s-T transformation, is a powerful
solution generating technique in string and supergravity theories
\cite{Dowker:1993bt,Dowker:1994up,Behrndt:1995si,Costa:2000nw,Gutperle:2001mb,Costa:2001ifa}.
Melvin twist of flat spaces retains the simplicity of the flat space,
giving rise to string theory whose world sheet theory is exactly
solvable
\cite{Russo:1994cv,Russo:1995tj,Russo:1995aj,Russo:1995ik,Tseytlin:1994ei,Tseytlin:1995fh}. Melvin
twists, in a context of D-branes and their near horizon limits, can be
used to formulate wide variety of exotic field theories, including
non-commutative gauge theories
\cite{Hashimoto:1999ut,Aharony:2000gz,Hashimoto:2002nr,Dolan:2002px,Hashimoto:2004pb,Hashimoto:2005hy,Dhokarh:2008ki},
NCOS theories
\cite{Seiberg:2000ms,Gopakumar:2000na,Barbon:2000sg,Cai:2002sv,Cai:2006tda},
dipole theories \cite{Bergman:2000cw,Bergman:2001rw,Ganor:2002ju}, and
$\beta$ deformed superconformal theories \cite{Lunin:2005jy}.

Recently, ``Puff Field Theory,'' a new class of decoupled non-local
field theory based on Melvin universe, was introduced by Ganor
\cite{Ganor:2006ub}. In the construction of PFT, the Melvin background
is supported by RR field strength, but the decoupled theory is
distinct\footnote{See appendix A of \cite{Ganor:2007qh} for a
discussion of this point.} from the NCOS theories
\cite{Cai:2002sv,Cai:2006tda}. In 3+1 dimensions, PFT preserves the
spatial $SO(3)$ subgroup of the $SO(1,3)$ Lorentz symmetry. The dual
supergravity formulation of PFT, along the lines of
\cite{Hashimoto:1999ut}, was constructed in \cite{Ganor:2007qh},
allowing physical feature of this model, such as the entropy as a
function of temperature, to be computed at large 't Hooft
coupling. One of the main appeal of PFT is the fact that it is
compatible with the symmetries of Freedman-Robertson-Walker
cosmology. Some phenomenological aspects of PFT were studied recently
in \cite{Minton:2007fd}.

The Melvin deformed field theories enumerated earlier: non-commutative
field theory, NCOS, dipole field theories, and $\beta$-deformed
superconformal field theories all have concrete formulations
independent of string theory. In contrast, only definition available
for PFT, for the time being, is as a decoupling limit of fluctuations
of D-branes in a Melvin geometry in type II string theory. The goal of
this article is to provide an alternative definition of PFT which is
independent of string theory.  Our approach will closely parallel the
formulation of Little String Theory and (0,2) superconformal field
theory using deconstruction \cite{ArkaniHamed:2001ie}.

Let us begin by reviewing the construction of PFT as a decoupled
theory on a brane in string theory \cite{Ganor:2006ub,Ganor:2007qh}. A
convenient place to start is flat 9+1 geometry in type IIA theory,
with $N$ coincident D0-branes. Let us ignore the gravitational back
reaction of the D0-branes for the time being. The M-theory lift of the
IIA geometry is $R^{1,9} \times S_1$.  Let us parameterize this
geometry with a line-element of the form
\be ds^2 = -dt^2 + dr^2 + r^2 d \phi^2 + d \vec y^2 + dz^2 \ee
where $z$ is the M theory circle with periodicity $z \sim z + 2 \pi
R$, $R = g_s l_s$, and $\vec y$ is a vector in seven dimensions.  The
$r$, $\phi$ parameterize a plane spanned by the remaining two
coordinates in cylindrical coordinates.

Now consider performing a Melvin twist on $\phi$ with respect to shift
in $z$. This amounts to deforming the line element by the amount
$\eta$ so that
\be ds_{11}^2 = -dt^2 + dr^2 + r^2 (d \phi + \eta dz)^2 + d \vec y^2 + dz^2 \ . \label{twist} \ee
Reducing this to IIA gives rise to a Melvin geometry of the form
\be ds_{IIA}^2= (1 + \eta^2 r^2)^{1/2} \left( -dt^2 + dr^2 + {r^2 \over 1+\eta^2 r^2} d \phi^2 + d \vec y^2 \right)  \ee
along with some RR 1-form and a dilaton. Recalling that there were $N$
D0-branes in the background, perform a T-duality along 3 of the $y_i$
coordinates. The prescription of \cite{Ganor:2006ub,Ganor:2007qh} is
to send $\alpha' \rightarrow 0$ keeping $g_{YM3}^2$ and $\Delta^3 =
\eta \alpha'^2$ fixed. It is straight forward to reproduce the
supergravity background of \cite{Ganor:2007qh} by repeating this
procedure, but including the gravitational back reaction of the
D0-branes.

One issue which was not emphasized in the discussions of
\cite{Ganor:2006ub,Ganor:2007qh} is that one can just as easily
construct a generalization of PFT in $p+1$ dimensions for $0 \le p \le
5$ by changing the number of T-dualities one performs, and scaling to
keep $g_{YMp}^2 \sim g_s \alpha'^{(p-3)/2}$ to stay finite in the
scaling limit. The resulting $p+1$ dimensional field theory will
preserve the $SO(p)$ subgroup of the Lorentz group.

To demonstrate the decoupling limit of PFT for general $p$ more
concretely, let us work out the supergravity dual of the $p=0$ case
explicitly.  With the gravitational back reaction of D0 taken into
account, (\ref{twist}) becomes
\be 
ds^2_{11} =  - h^{-1} dt^2  +   h (d z - v dt)^2  + d \rho^2  + \rho^2  
(ds_{B(2)}^2 +  (d \phi  + \eta d  z + {\cal A})^2 )
 + \sum_{i = 1}^{5} d y_i^2  \label{mlift}
\ee
where
\be 
h(\rho,y) = 1 +  {g N \alpha'^{7/2} \over (\rho^2 + \vec y^2)^{7/2}} , \qquad v = h^{-1}
\ee
is the harmonic function of a D0-brane, and
\be d\Omega_3^2 = ds^2_{B(2)} + (d \phi  + {\cal A})^2 \ee
is the standard Hopf parameterization of $S^3$ with the $B(2)$ being
the base $S^2$. In writing this geometry, we generalized the twist
from being along the angular coordinate $\phi$ in a plane in
(\ref{twist}), to being along the Hopf fiber of angular 3-sphere in
$R^4$ spanned by {\it four} of the $\vec y$ coordinates.  This change
essentially amounts to considering the F5 flux brane instead of F7
flux brane in the language of \cite{Gutperle:2001mb}.  The latter
choice has the advantage of preserving half of the supersymmetries.
Now, reduce to IIA and take a decoupling limit, by scaling $\alpha'
\rightarrow 0$ keeping $U = r/\alpha'$, $\Delta^3 = \eta \alpha'^2$,
and $g_{YM0}^2 = g_s \alpha'^{-3/2}$ fixed. Note that
\be \nu = g_{YM0}^2 \Delta^3 \ee
is a dimensionless and a finite quantity. This parameter will play an
important role in the discussions below.

After taking the $\alpha' \rightarrow 0$ limit, we arrive at a solution
of type IIA supergravity of the form
\beq {ds^2\over \alpha'}  &=&  \sqrt{H + \Delta^6 U^2} \left( - H^{-1} dt^2   + d
U^2  + U^2
 d s_{B(2)}^2 + U^2\left(d \phi + {\cal A}+ {\Delta^3 \over H} dt\right)^2 +d\vec Y^2 \right) \cr
{A \over \alpha'{}^2} & = &  {1 \over H  + \Delta^6 U^2} \left( - dt + U^2 \Delta^3 d \phi \right) \label{pqm} \\
e^\phi & = & g_{YM}^2 (H + \Delta^6 U^2)^{3/4} \nonumber
\eeq
where
\be U = {\rho \over \alpha'}, \quad  \vec Y = {y \over \alpha'},\quad  H(U,\vec Y) = \alpha'^2 h(\rho,\vec y) =
{g_{YM0}^2 N \over  (U^2+Y^2)^{7/2}} \ 
\ee
have  finite  $\alpha' \rightarrow 0$ limits.

This geometry has a natural form to correspond to a supergravity dual
of decoupled theory on D0-branes. It is straight forward to generalize
this construction to other values of $p$.

Let us refer to the decoupled theory for $p=0$ as ``Puff Quantum
Mechanics.''  If we set $\Delta=0$, the metric (\ref{pqm}) precisely
reduces to the near horizon limit of D0-branes
\cite{Itzhaki:1998dd}. From the form of (\ref{pqm}), one can infer
that $\Delta$ deforms the matrix quantum mechanics of the decoupled
D0-branes in the UV. It is also clear that the dynamics of the
decoupled theory is somehow being modified by the RR 1-form potential
in the background. In the remainder of this article, we will provide a
prescription to define PQM independent of string theory.

One powerful tool in analyzing microscopic features of non-local field
theories is the $SL(2,Z)$ duality. In the case of non-commutative
field theory on a torus, an $SL(2,Z)$ duality is realized in the form
of Morita equivalence \cite{Pioline:1999xg}. If the deformation
parameter, e.g. the non-commutativity parameter $2 \pi \Delta^2$, is
expressed in a suitably dimensionless form, e.g. $\Theta = \Delta^2
/\mbox{Vol}(T^2)$, then for a rational value of $\Theta$, one can find
an $SL(2,Z)$ element to map this theory to a dual theory for which
$\Theta=0$. For non-commutative field theories, the $\Theta=0$ theory
corresponds to the standard non-abelian gauge theory on $T^2$ with a
't Hooft flux \cite{tHooft:1981sz}. Various $SL(2,Z)$ duals have
non-overlapping regimes of validity as a function of energy, giving
rise to a structure resembling a duality cascade
\cite{Hashimoto:1999yj}. For all rational values of $\Theta$, it is
the $SL(2,Z)$ dual with vanishing $\Theta$ which is effective in the
deep UV.

Analogous $SL(2,Z)$ structure exists for PFT \cite{Ganor:2007qh}. In
the context of PQM, this structure is made most transparent by
performing a modular transformation on the complex structure of the
torus defined by $\phi$ and $z$ in (\ref{mlift}), i.e.
\be \left(\begin{array}{c} d \phi  \\ {d z \over  R} \end{array} \right) 
\rightarrow \left( \begin{array}{cc} a & b \\ c & d \end{array}\right)
\left(\begin{array}{c} d \phi  \\ {d z \over  R} \end{array} \right) 
\ee
If the dimensionless parameter
\be \nu = g_{YM0}^2 \Delta^3 = \eta R=  -{b \over d} \ee
is a rational number, this $SL(2,Z)$ will map (\ref{mlift}) to 
\beq 
ds^2 & =&  -  h^{-1/2} dt^2  +  h^{1/2}\left(  d \rho^2  + \rho^2  
ds_{B(2)}^2 +   \rho^2\left({d \phi \over d} + {\cal A} \right)^2
 +d\vec y^2\right) \cr
A & = & -  c   R d \phi - v   dt \cr
e^{\phi} & = &  h^{3/4}\ .  \nonumber
\eeq

Other than the seemingly innocent 1-form $A = c  R d \phi$, this
is just a $Z_d$ orbifold of D0 which defines a local theory in the
decoupling limit. The $SL(2,Z)$ also changed the rank of the gauge
group from $U(N)$ to $U(d^2 N)/Z_d$, as well as acting on $g_{YM0}^2$.

Let us refer to the $RR$ deformed D0-brane quantum mechanics as
``Twist Quiver Quantum Mechanics.''  Our goal is to determine how the
$RR$ background affects the dynamics of TQQM.

\begin{figure}
\centerline{\includegraphics[width=2.3in]{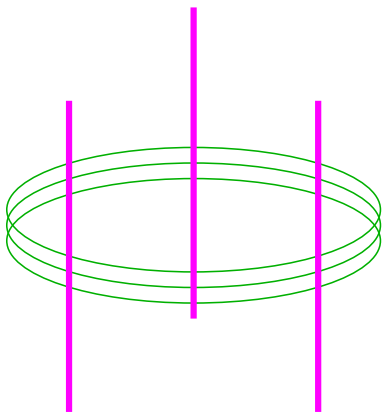}}
\caption{Configuration of $dN$ D1-brane and $d$ NS5 impurities obtained by T-dualizing the background (\ref{mlift}) along the $\phi$ direction. There is also a RR axion $\chi={c \over d}$ in this background.\label{figa}}
\end{figure}

In order to access the effect of form fields on the dynamics of open
string states, it is convenient to go to a dual frame where the form
field in question is mapped to an NSNS 2-form so that one has access
to an explicit NSR sigma model. In order to accomplish this, let us
momentarily embed the $Z_d$ ALE orbifold spanned by $\rho$, $B(2)$,
and $\phi$ in a Taub-NUT. This is a UV modification which can be
removed later. The reason for embedding into the Taub-NUT is to
facilitate the T-duality along $\phi$. This T-duality will map the
Taub-NUT to NS5-branes, and the D0-brane to a D1-brane. They are oriented as follows:
\centerline{\begin{tabular}{lllllllllll}
   & 0 & 1 & 2 & 3 & 4 & 5 & 6 & 7 & 8 & 9 \cr
D1 & \bl & \bl  \cr
NS5 & \bl &  & \bl& \bl& \bl& \bl& \bl
\end{tabular}}
The system
is to be visualized as a system of $dN$ D1-branes sprinkled with $d$ NS5
impurities, which is illustrated in figure \ref{figa}.  The RR 1-form
potential $A = {c \over d} (d  R) d \phi$ becomes the RR axion $\chi = {c \over d}$ under this duality. 

In order to map the RR axion into NSNS 2-form, we further compactify
two directions, parallel to the NS5-brane world volume but orthogonal
to the D1. This compactification will also deform the theory in the
UV, which we will remove at the end of the construction. T-dualizing
along these two directions, followed by S-duality will lead to a
system consisting of $dN$ D3-branes on $T^2$, with $d$ D5 impurities
extended along the $T^2$ and localized in the remaining spatial
coordinate of the D3, in a background a constant NSNS 2-form along the
$T^2$.
\centerline{\begin{tabular}{lllllllllll}
   & 0 & 1 & 2 & 3 & 4 & 5 & 6 & 7 & 8 & 9 \cr
$B$-field & & & $\equiv$ & $\equiv$   \cr
D3 & \bl & \bl& \bl& \bl  \cr
D5 & \bl &  & \bl& \bl& \bl& \bl& \bl
\end{tabular}}
Except for the compactification and the $B$-fields, this is precisely
the impurity model of Karch and Randall \cite{Karch:2001cw} whose
detailed microscopic formulation was given in \cite{DeWolfe:2001pq}.

Of course, to isolate the PQM/TQQM dynamics, we are only interested in
the deep IR where only the dimensionally reduced dynamics matters. We
can take advantage of this fact to further reformulate this system by
performing additional dualities.

\begin{figure}
\centerline{\begin{tabular}{cc}
\qquad\qquad\qquad\includegraphics{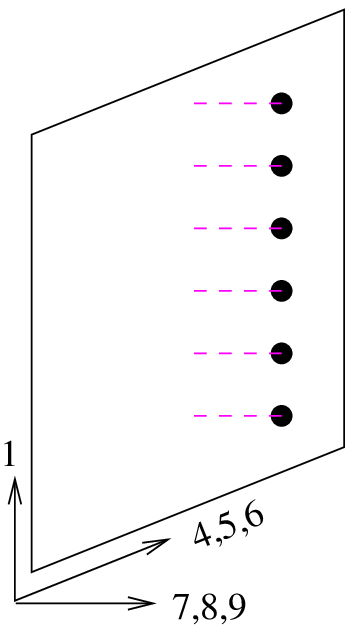}\qquad\qquad\qquad & \qquad\qquad\qquad\includegraphics{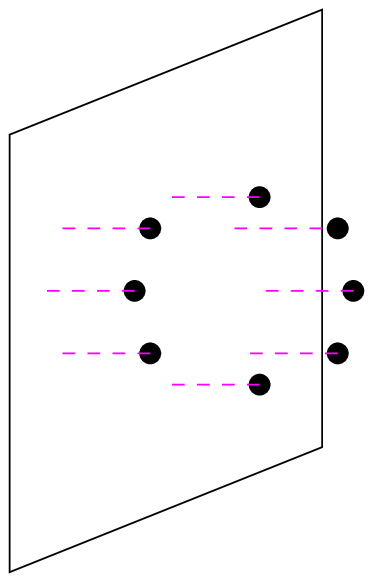} \qquad\qquad\qquad \\
(a) & (b) 
\end{tabular}} 
\caption{(a) T-dual of Defect Field Theory. The D2-branes is localized
in a circle. In the covering space, it corresponds to a $U(\infty)/Z$
theory corresponding to an array of D2-branes in a background of
D6-branes. (b) The infinite array can be arrived as a limit of a
circular lattice by scaling the radius and the number of branes
keeping the linear density of the branes fixed.\label{figb}. For the
sake of illusteration, the D2-branes appear separated from the
D6-branes along the 7,8,9 directions in the figure. The actual
configuration we consider, the D2 and the D6 branes are coincident
along the 7,8,9 directions.}
\end{figure}

Consider T-dualizing along the world volume of the D3-brane in the 
$x_1$ direction  along which the D1 was originally
oriented. This will map the D3-brane to a D2-brane localized in a
circle. Its covering space is an infinite array.  The impurity
D5-branes are mapped to an extended D6. The $T^2$ along which the
$B$-field was oriented is left intact. 
\centerline{\begin{tabular}{lllllllllll}
   & 0 & 1 & 2 & 3 & 4 & 5 & 6 & 7 & 8 & 9 \cr
$B$-field & & & $\equiv$ & $\equiv$   \cr
D2 & \bl & & \bl& \bl  \cr
D6 & \bl & \bl & \bl& \bl& \bl& \bl& \bl
\end{tabular}}
This configuration is illustrated in 
figure \ref{figb}.a. What we have done is to exchange the Kaluza-Klein mode associated with the $x_1$ direction to a tower of massive $W$ bosons in a $U(\infty) / Z$ gauge theory along the lines of \cite{Taylor:1996ik}.

Let us now employ a trick of presenting $U(\infty)/Z$ as a limit of
$U(N)/Z_N$. Simply consider arranging $N$ D2 in a circular, instead of
the linear, pattern as is illustrated in figure \ref{figb}.b. This is
essentially a technique to simulate T-duality via deconstruction along
the lines of \cite{ArkaniHamed:2001ie}.
				   
What we have now is a configuration of D2 in Coulomb branch, in a
background of D6-branes. This is essentially the configuration which
gives rise to 2+1 SYM with flavor, considered in
\cite{Cherkis:2002ir}. The only novelty here is the fact that the
world volume of D2 is compactified on a torus, and that there is a
$B$-field oriented along it.

Following the duality chain, it should be clear that each dot in
figure \ref{figb} should correspond to $dN$ D2-branes.  It is quite
natural therefore to interpret the $c/d$ units of $B$-flux as giving
rise to 't Hooft's fractional flux on the D2-brane world volume gauge
theory
\cite{tHooft:1981sz,vanBaal:1982ag,Guralnik:1997sy,Hashimoto:1997gm}.

There is one subtlety with this interpretation. In order for the 't
Hooft flux to exist as a consistent field configuration, it is
necessary for all the fields in the theory to be invariant with
respect to the center of the gauge group. The flavor matter which
arise in our setup due to the presence of the D6-brane, is in the
fundamental representation with respect to the gauge group and does
not satisfy this requirement.

A moment's thought, however, suffices to address this issue. One
simply needs to recall that the 2-6 strings are actually in a
bifundamental representation of the D2 and the D6-brane gauge fields.
Since there are $d$ D6-branes in the configuration illustrated in
figure \ref{figb}, it can also support a flux in the amount of $\int
B=c/d$. From the point of view of the 2+1 dimensional Yang-Mills
theory, this amounts to twisting with respect to gauge and the flavor
group of the bifundamentals.  In other words, we parameterize the
color and flavor indices of the bi-fundamental as $\Phi_{i,c;f}$
where $i = 1 \ldots N$, $c=1 \ldots d$, and $f = 1 \ldots d$. As the
notation suggests, $(i,c)$ are the color indices and $f$ is the
flavor index. To these bi-fundamental fields, we impose the boundary
condition
\beq \Phi_{i,c;f} (x + L_2,x_3) &=& U_{c,c'}\Phi_{i,c';f'} (x_2 ,x_3)
U^{-1}_{f',f} \cr \Phi_{i,c;f} (x_2,x_3 + L_3) &=&
V_{c,c'}\Phi_{i,c';f'} (x_2,x_3) V^{-1}_{f',f} \label{bc} \eeq 
where $U$ and $V$ are $d\times d$ 't Hooft matrices
\cite{tHooft:1981sz} satisfying
\be U V U^{-1} V^{-1} = e^{2 \pi i c/d} \ . \ee
In fact, precisely this form of twisted matter theory have been used
before by Sumit Das in the context of large $N$ twisted reduced models
\cite{Das:1983pm}. See also  \cite{Eguchi:1982nm,GonzalezArroyo:1982ub,Levine:1982uz,Cohen:1983sd} for discussons on related issues.

Of course, since we have performed various UV deformation of the
original PFT to get to this stage, one must take the appropriate
scaling limit to decouple these effects.  The fact that the decoupled
supergravity dual solution (\ref{pqm}) exists provides us with the
assurance that such a limit does exist. Because the chain of duality
involved S-duality at one point, what we are doing is similar in
spirit to defining NCOS as the strong coupling limit of NCSYM
\cite{Seiberg:2000ms,Gopakumar:2000na,Barbon:2000sg}.

In summary, we have shown that PQM is a scaling limit, of a large $N$
deconstruction limit, of 2+1 dimensional SYM, with 't Hooft flux, and
matter in the fundamental representation, with twisted flavor. The
derivation relied on a lengthy chain of dualities and manipulations in
string theory. Nonetheless, the formulation of the theory in its final
form does not rely on any string theory concepts. While this
definition is not especially useful for most practical applications,
it does provide a concrete formulation of the theory whose only other
formulation known today is as a decoupling limit of D-branes in a
Melvin universe background.  \cite{Ganor:2006ub,Ganor:2007qh}.

The color/flavor twisted 2+1 dimensional theory (\ref{bc}) might be an
interesting theory in its own right to explore further.  These
theories are related via Morita equivalence to non-commutative field
theories with matter \cite{Ambjorn:2000nb,Ambjorn:2000cs}, whose dual
supergravity solution was briefly described in
\cite{Cherkis:2002ir}. It might also be interesting to explore how the
twists and non-commutativities modify the Intriligator Seiberg mirror
symmetry of three dimensional gauge theories
\cite{Intriligator:1996ex}.

\section*{Acknowledgements}
We would like to thank
Sharon Jue, Bom Soo Kim, Anthony Ndirango, and especially Ori Ganor
for collaboration which motivated this project, and for useful
discussions. We also thank Sumit Das for bringing his early work on
large $N$ twist reduced models to our attention.  This work was
supported in part by the DOE grant DE-FG02-95ER40896 and funds from
the University of Wisconsin.

\providecommand{\href}[2]{#2}\begingroup\raggedright\endgroup

\end{document}